%Paper: hep-th/9309125
%From: Chris Pope <pope@avignon.tamu.edu>
%Date: Wed, 22 Sep 93 22:21:28 -0600
%Date (revised): Wed, 22 Sep 93 22:33:38 -0600
%Date (revised): Wed, 22 Sep 93 22:37:03 -0600

%%%%%%%%%%%%%%%%%%%%%%%%%%%%%%%%%%%%%%%%%%%%%%%%%%%%%%%%%%%%%
%%%                                                       %%%
%%%                   $W$-Strings 93                      %%%
%%%                                                       %%%
%%%                     C.N. Pope                         %%%
%%%                                                       %%%
%%%                    USE PLAIN TEX                      %%%
%%%                                                       %%%
%%%%%%%%%%%%%%%%%%%%%%%%%%%%%%%%%%%%%%%%%%%%%%%%%%%%%%%%%%%%%
%\input manumac
\def\singlespace{\normalbaselines}
\def\oneandahalfspace{\baselineskip=1.15\normalbaselineskip plus 1pt
\lineskip=2pt\lineskiplimit=1pt}

\def\np{\vfill\eject}
\def\nl{\hfil\break}

\def\nofirstpagenoten{\nopagenumbers\footline={\ifnum\pageno>1\tenrm
\hss\folio\hss\fi}}
\def\nofirstpagenotwelve{\nopagenumbers\footline={\ifnum\pageno>1\twelverm
\hss\folio\hss\fi}}
\def\leaderfill{\leaders\hbox to 1em{\hss.\hss}\hfill}
\def\ft#1#2{{\textstyle{{#1}\over{#2}}}}
\def\frac#1/#2{\leavevmode\kern.1em
\raise.5ex\hbox{\the\scriptfont0 #1}\kern-.1em/\kern-.15em
\lower.25ex\hbox{\the\scriptfont0 #2}}
\def\sfrac#1/#2{\leavevmode\kern.1em
\raise.5ex\hbox{\the\scriptscriptfont0 #1}\kern-.1em/\kern-.15em
\lower.25ex\hbox{\the\scriptscriptfont0 #2}}

  %20 point
                   %17 point
  %14 point
 %17 point
 %14 point
 %14 point
 %14 point

\parindent=20pt
\def\narrow{\advance\leftskip by 40pt \advance\rightskip by 40pt}

\def\AB{\bigskip
        \centerline{\bf ABSTRACT}\medskip\narrow}
\def\nonarrower{\advance\leftskip by -40pt\advance\rightskip by -40pt}
\def\AE{\bigskip\nonarrower}

\def\boxit#1{\vbox{\hrule\hbox{\vrule\kern3pt
        \vbox{\kern3pt#1\kern3pt}\kern3pt\vrule}\hrule}}

\def\gtorder{\mathrel{\raise.3ex\hbox{$>$}\mkern-14mu
             \lower0.6ex\hbox{$\sim$}}}
\def\ltorder{\mathrel{\raise.3ex\hbox{$<$}|mkern-14mu
             \lower0.6ex\hbox{\sim$}}}
\def\dalemb#1#2{{\vbox{\hrule height .#2pt
        \hbox{\vrule width.#2pt height#1pt \kern#1pt
                \vrule width.#2pt}
        \hrule height.#2pt}}}

\font\fourteentt=cmtt10 scaled \magstep2
\font\fourteenbf=cmbx12 scaled \magstep1
\font\fourteenrm=cmr12 scaled \magstep1
\font\fourteeni=cmmi12 scaled \magstep1
\font\fourteenss=cmss12 scaled \magstep1
\font\fourteensy=cmsy10 scaled \magstep2
\font\fourteensl=cmsl12 scaled \magstep1
\font\fourteenex=cmex10 scaled \magstep2
\font\fourteenit=cmti12 scaled \magstep1
\font\twelvett=cmtt10 scaled \magstep1 \font\twelvebf=cmbx12
\font\twelverm=cmr12 \font\twelvei=cmmi12
\font\twelvess=cmss12 \font\twelvesy=cmsy10 scaled \magstep1
\font\twelvesl=cmsl12 \font\twelveex=cmex10 scaled \magstep1
\font\twelveit=cmti12
\font\tenss=cmss10
 
 \font\ninebf=cmbx7 scaled \magstep1
\font\ninerm=cmr7 scaled \magstep1 \font\ninei=cmmi7 scaled \magstep1
\font\ninesy=cmsy7 scaled \magstep1 
\font\eightrm=cmr7 scaled 1140 
 
\font\sevenbf=cmbx7 \font\sevenrm=cmr7 \font\seveni=cmmi7
\font\sevensy=cmsy7 

\catcode`@=11
\newskip\ttglue
\newfam\ssfam

\def\fourteenpoint{\def\rm{\fam0\fourteenrm}
\textfont0=\fourteenrm \scriptfont0=\tenrm \scriptscriptfont0=\sevenrm
\textfont1=\fourteeni \scriptfont1=\teni \scriptscriptfont1=\seveni
\textfont2=\fourteensy \scriptfont2=\tensy \scriptscriptfont2=\sevensy
\textfont3=\fourteenex \scriptfont3=\fourteenex \scriptscriptfont3=\fourteenex
\def\it{\fam\itfam\fourteenit} \textfont\itfam=\fourteenit
\def\sl{\fam\slfam\fourteensl} \textfont\slfam=\fourteensl
\def\bf{\fam\bffam\fourteenbf} \textfont\bffam=\fourteenbf
\scriptfont\bffam=\tenbf \scriptscriptfont\bffam=\sevenbf
\def\tt{\fam\ttfam\fourteentt} \textfont\ttfam=\fourteentt
\def\ss{\fam\ssfam\fourteenss} \textfont\ssfam=\fourteenss
\tt \ttglue=.5em plus .25em minus .15em
\normalbaselineskip=16pt
\abovedisplayskip=16pt plus 4pt minus 12pt
\belowdisplayskip=16pt plus 4pt minus 12pt
\abovedisplayshortskip=0pt plus 4pt
\belowdisplayshortskip=9pt plus 4pt minus 6pt
\parskip=5pt plus 1.5pt
\setbox\strutbox=\hbox{\vrule height12pt depth5pt width0pt}
\let\sc=\tenrm
\let\big=\fourteenbig \normalbaselines\rm}
\def\fourteenbig#1{{\hbox{$\left#1\vbox to12pt{}\right.\n@space$}}}

\def\twelvepoint{\def\rm{\fam0\twelverm}
\textfont0=\twelverm \scriptfont0=\ninerm \scriptscriptfont0=\sevenrm
\textfont1=\twelvei \scriptfont1=\ninei \scriptscriptfont1=\seveni
\textfont2=\twelvesy \scriptfont2=\ninesy \scriptscriptfont2=\sevensy
\textfont3=\twelveex \scriptfont3=\twelveex \scriptscriptfont3=\twelveex
\def\it{\fam\itfam\twelveit} \textfont\itfam=\twelveit
\def\sl{\fam\slfam\twelvesl} \textfont\slfam=\twelvesl
\def\bf{\fam\bffam\twelvebf} \textfont\bffam=\twelvebf
\scriptfont\bffam=\ninebf \scriptscriptfont\bffam=\sevenbf
\def\tt{\fam\ttfam\twelvett} \textfont\ttfam=\twelvett
\def\ss{\fam\ssfam\twelvess} \textfont\ssfam=\twelvess
\tt \ttglue=.5em plus .25em minus .15em
\normalbaselineskip=14pt
\abovedisplayskip=14pt plus 3pt minus 10pt
\belowdisplayskip=14pt plus 3pt minus 10pt
\abovedisplayshortskip=0pt plus 3pt
\belowdisplayshortskip=8pt plus 3pt minus 5pt
\parskip=3pt plus 1.5pt
\setbox\strutbox=\hbox{\vrule height10pt depth4pt width0pt}
\let\sc=\ninerm
\let\big=\twelvebig \normalbaselines\rm}
\def\twelvebig#1{{\hbox{$\left#1\vbox to10pt{}\right.\n@space$}}}

\def\tenpoint{\def\rm{\fam0\tenrm}
\textfont0=\tenrm \scriptfont0=\sevenrm \scriptscriptfont0=\fiverm
\textfont1=\teni \scriptfont1=\seveni \scriptscriptfont1=\fivei
\textfont2=\tensy \scriptfont2=\sevensy \scriptscriptfont2=\fivesy
\textfont3=\tenex \scriptfont3=\tenex \scriptscriptfont3=\tenex
\def\it{\fam\itfam\tenit} \textfont\itfam=\tenit
\def\sl{\fam\slfam\tensl} \textfont\slfam=\tensl
\def\bf{\fam\bffam\tenbf} \textfont\bffam=\tenbf
\scriptfont\bffam=\sevenbf \scriptscriptfont\bffam=\fivebf
\def\tt{\fam\ttfam\tentt} \textfont\ttfam=\tentt
\def\ss{\fam\ssfam\tenss} \textfont\ssfam=\tenss
\tt \ttglue=.5em plus .25em minus .15em
\normalbaselineskip=12pt
\abovedisplayskip=12pt plus 3pt minus 9pt
\belowdisplayskip=12pt plus 3pt minus 9pt
\abovedisplayshortskip=0pt plus 3pt
\belowdisplayshortskip=7pt plus 3pt minus 4pt
\parskip=0.0pt plus 1.0pt
\setbox\strutbox=\hbox{\vrule height8.5pt depth3.5pt width0pt}
\let\sc=\eightrm
\let\big=\tenbig \normalbaselines\rm}
\def\tenbig#1{{\hbox{$\left#1\vbox to8.5pt{}\right.\n@space$}}}
\let\rawfootnote=\footnote \def\footnote#1#2{{\rm\parskip=0pt\rawfootnote{#1}
{#2\hfill\vrule height 0pt depth 6pt width 0pt}}}

\def\tenfoot{\tenpoint\hskip-\parindent\hskip-.1cm}

\overfullrule=0pt
\twelvepoint
\def\sbullet{\raise.2em\hbox{$\scriptscriptstyle\bullet$}}
\nofirstpagenotwelve
%\magnification=\magstep1%
\hsize=16.5 truecm
%\vsize=23.0 truecm
\baselineskip 15pt
%\parskip 0pt
%\font\ftf=cmr8

\def\ft#1#2{{\textstyle{{#1}\over{#2}}}}
\def\sss{\scriptscriptstyle}
\def\ket#1{\big| #1\big\rangle}

\def\fpf#1#2#3#4{\big\langle #1\ #2\ #3\ #4\big\rangle}
\def\fipf#1#2#3#4#5{\big\langle #1\ #2\ #3\ #4\ #5\big\rangle}

\def\teff{T^{\rm eff}}

\def\del{\partial}

\def\cramp{\medmuskip = -3mu plus 1mu minus 2mu}
\def\uncramp{\medmuskip = 4mu plus 2mu minus 4mu}

\oneandahalfspace
\rightline{CTP TAMU--55/93}
\rightline{hep-th/9309125}
\rightline{September 1993}

\vskip 2truecm
\centerline{\bf $W$-Strings 93\footnote{$^\dagger$}{\tenfoot Based on Talks
presented at the Spring Workshop on High-energy Physics, Trieste, April 1993,
and \nl
\phantom{xxxxx}{\it Strings '93}, Berkeley, May 1993.}}
\vskip 1truecm
\centerline{C.N. Pope\footnote{$^*$}{\tenfoot Supported in part by the U.S.
Department of Energy, under grant DE-FG05-91ER40633.}}
\vskip 1.5truecm
\centerline{\it Center
for Theoretical Physics,
Texas A\&M University,}
\centerline{\it College Station, TX 77843--4242, USA.}

\vskip 2truecm

\AB\singlespace
   We present a review of the status of $W$ string theories, their
physical spectra, and their interactions.
\AE\oneandahalfspace

%\bigskip\bigskip
\np

\noindent{\bf 1. Introduction}
\medskip

      $W$ algebras have received a considerable amount of attention since
their discovery [1] in the middle of the 1980's.  These efforts have broadly
concentrated in two directions, namely the construction and classification of
$W$ algebras on the one hand, and their application to $W$ gravity and $W$
strings on the other.   Since $W$ algebras are higher-spin extensions of the
Virasoro algebra, it is natural to expect that they could be associated with
some kind of extensions of the usual bosonic string.  The idea of building
$W$-string theories was first developed in [2].  One of the hopes was that they
might have massless physical states with spins greater than 2 [2], but so far
all the explicit examples that have been constructed do not give rise to such
higher-spin massless states.  Nonetheless, the structures that emerge in
$W$-string theories are not without interest, revealing intriguing connections
with Virasoro and $W$ minimal models.

     Most of the efforts so far in constructing $W$-string theories have been
concentrated on the case of the $W_3$ string, since $W_3$ is the simplest
non-trivial $W$ algebra.  It has a primary current of spin 3 in addition to the
energy-momentum tensor.   Like all $W$ algebras, it is non-linear.  The
starting point for building a $W$ string is to construct an anomaly-free
quantum theory of the associated $W$ gravity.  By far the
simplest way to do this is by BRST methods [3].  The essential requirement,
therefore, is to find the BRST operator $Q_B$ for the $W$ algebra in question.
{}From this, one can write down the full Lagrangian, with ghosts and
gauge-fixing
terms.  The requirement of anomaly-freedom translates into the requirement that
the BRST operator $Q_B$ be nilpotent.  This is achieved provided that the
matter
currents satisfy a closed (albeit non-linear) algebra  at the full quantum
level, with the correct value of central charge to cancel the anomaly from
the ghosts for the gauge fields.

     Having arrived at an anomaly-free quantum $W$ gravity, the next step in
building a $W$ string is to find an appropriate $W$ matter system, with the
critical value of the central charge, that admits some sort of a spacetime
interpretation.  Thus some of the matter fields,
which we may suggestively call $X^\mu$, should be worldsheet scalars that enter
the matter currents in a Lorentz-invariant way.  It is non-trivial that this
can be done, since the $W$ algebra is non-linear and one  cannot, by contrast
with ordinary string theory, simply take tensor products of elementary
realisations to obtain new ones with larger central charge.

     The basic elementary realisations of $W$ algebras are those that come from
Hamiltonian reduction and the Miura transformation.  In the case of $W_3$,
for example, this is a two-scalar realisation [1].  The central charge can be
adjusted to its critical value ($c=100$ for $W_3$) by an appropriate choice of
a free background-charge parameter.  This realisation gives rise to what one
may call a theory of pure $W_3$ gravity, and can be thought of as a
generalisation of a one-scalar realisation of the Virasoro algebra, or pure
Liouville gravity.

     In [4], the observation was made that one of the two-scalars of the Miura
realisation of $W_3$ appears in the spin-2 and spin-3 currents only through
its energy-momentum tensor.  Thus one can obtain more general realisations
of $W_3$ by replacing this energy-momentum tensor by that for an arbitrary
matter system $T^{\rm eff}$ with the same central charge $c^{\rm eff}$, the
only other requirement being that it should commute with the other original
scalar field, which we shall call $\varphi$.  The value of $c^{\rm eff}$,
which is dictated by the details of the Miura realisation and the required
critical value ($c=100$) for the total matter central charge, turns out to
be $c^{\rm eff}=\ft{51}2$.  It is no coincidence, as we shall see later,
that this is 26 minus the central charge of the Ising model.  By realising
$T^{\rm eff}$ in terms of worldsheet scalars $X^\mu$, the desired goal of
obtaining a $W_3$ realisation with a string-like spacetime interpretation is
achieved.  The first investigations of the $W_3$ string based on this
realisation were carried out in [5].

     Although these multi-scalar realisations enable
one to construct $W_3$ strings that admit a sensible-looking multi-dimensional
spacetime interpretation, in some sense the way in which the $W_3$
symmetry is realised on the fields $X^\mu$ is a little trivial, and this
ultimately shows up when one calculates the spectrum and scattering of physical
states in the theory.  As we shall see later, the multi-scalar $W_3$ string
behaves, at least at tree level, precisely like a bosonic string in which
criticality is achieved by tensoring a $c^{\rm eff}=\ft{51}2$ energy-momentum
tensor for $X^\mu$ (with a background charge) with the $c=\ft12$ Ising model.
While it is not without interest that such a system displays a ``hidden'' $W_3$
symmetry, the original hopes that $W$ strings might be new kinds of
string theories, maybe even with higher-spin massless states, seem to have
evaporated.

     In fact, although it is possibly physically less realistic, the
two-dimensional $W_3$ string based on the original two-scalar Miura realisation
is mathematically-speaking a much more interesting object.  It turns out that
the physical states of the two-scalar $W_3$ string divide into two categories;
there are some which generalise to the multi-scalar case if one makes the
replacement described above, and then there are other physical states that have
no analogues in the multi-scalar case.  The first category is in some sense
``uninteresting,'' in that again the physical states are tensor products of
$c^{\rm eff}=\ft{51}2$ Virasoro states with Ising-model primary fields.  The
second category comprises physical states that are not simply tensor products
of Virasoro-string states with Ising model states, and as such they could be
regarded as states on which the $W_3$ symmetry is realised in a more
non-trivial way.

     One could, of course, build a $W$ string theory based on more or less any
$W$ algebra, such as $W_N$.  However, since each higher-spin current in a $W$
algebra represents a constraint on the physical states of the associated $W$
string, there is a sense in which one obtains richer $W$-string theories by
considering algebras with fewer higher-spin currents.  To this end, recently
some examples of other kinds of $W$ strings have been constructed, in which
there is just a single current of spin $s>2$ in addition to the energy-momentum
tensor [6].  Curiously, it has been found that one does not necessarily even
need an underlying $W$ algebra in  order to construct a nilpotent BRST
operator.   We shall describe some recent work on these spin-2 plus spin-$s$
strings later.

\np
%\bigskip
\noindent{\bf 2. $W_3$ strings}
\medskip
\noindent{\it 2.1 The physical spectrum}
\medskip

     We shall begin by describing the multi-scalar $W_3$ string.  As was
discussed in the introduction, the key ingredient for the construction of the
$W_3$ string is its BRST operator.  This was found in [7] (see also [8]).  It
takes the form
$$
Q_B=\oint dz \Big[c\,(T+\ft12 T_{\rm gh})+\gamma\,(W+\ft12
W_{\rm gh})\Big], \eqno(2.1)
$$
and is nilpotent provided that the matter currents $T$ and $W$ generate the
$W_3$ algebra with central charge $c=100$, and that the ghost currents are
chosen  to be
$$
\cramp
\eqalignno{
T_{\rm gh}&=-2b\,\partial c-\partial
b\, c-3\beta\, \partial\gamma-2\partial\beta\, \gamma\ , &(2.2)\cr
W_{\rm gh}&=-\partial\beta\,
c-3\beta\, \partial c-\ft8{261}\big[\partial(b\, \gamma\,  T)+b\,
\partial\gamma \, T\big]
+\ft{25}{1566}\Big(2\gamma\, \partial^3b+9\partial\gamma\,
\partial^2b +15\partial^2\gamma\,\partial b+10\partial^3\gamma\,
b\Big) ,&(2.3)\cr}
\uncramp
$$
where the ghost-antighost pairs ($c$, $b$) and ($\gamma$, $\beta$) correspond
respectively to the $T$ and $W$ generators. A matter
realisation of $W_3$ with central charge 100 can be given
in terms of $n\ge 2$ scalar fields, as follows [4]:
$$
\eqalign{
T&= -\ft12 (\del\varphi)^2 - \alpha\, \del^2 \varphi +\teff,\cr
W&=-{2i \over \sqrt{261} }\Big[ \ft13 (\del\varphi)^3 + \alpha\, \del\varphi
\,\del^2\varphi +\ft13 \alpha^2 \,\del^3 \varphi + 2\,\del\varphi\, \teff
+ \alpha\, \del \teff\Big],\cr}\eqno(2.4)
$$
where $\alpha^2=\ft{49}{8}$ and $\teff$ is an energy-momentum tensor
with central charge $\ft{51}2$ that commutes with $\varphi$.  Since
$\teff$ has a fractional central charge, a background-charge vector
$a_\mu$  is needed in order to realise it with $d$ scalar fields $X^\mu$:
$$
\teff=-\ft12 \del X_\mu \del X^\mu - i a_\mu\, \del^2 X^\mu,\eqno(2.5)
$$
with $a_\mu$ chosen so that $\ft{51}2=d - 12 a_\mu a^\mu$.

     It is now in principle a straightforward matter to construct physical
states $\ket{\chi}$, by demanding that they be annihilated by the BRST
operator and that they be BRST non-trivial:
$$
Q_B\ket{\chi}=0,\qquad \ket{\chi}\ne Q_B\ket{\psi}. \eqno(2.6)
$$
Rather than follow the historical course of events here, we shall jump ahead
to a discovery made in [9], which leads to a considerable simplification of the
BRST operator and the form of the physical states.  In [9], the the following
non-linear redefinition, under which the ghosts and the $\varphi$ matter
field become mixed, was introduced:
$$
\eqalign{
c &\longrightarrow  c +\ft{21}{8 \sqrt2}\, \del\gamma
+\ft98\,  b\, \del \gamma\,  \gamma
-\ft32\, \del\varphi\, \gamma \ ,\cr
b &\longrightarrow  b\ ,\cr
\gamma &\longrightarrow \ft{9\sqrt{29}i}{8}\gamma\ ,\cr
\beta &\longrightarrow  -\ft{8i}{9\sqrt{29}}\Big\{\beta +
\ft{21}{ 8\sqrt2}\, \del  b  +\ft98\del  b\,  b\,  \gamma
+\ft32\, \del \varphi\,  b\ \Big\},\cr
\varphi &\longrightarrow \varphi-\ft32\,  b\,
 \gamma\ .\cr}\eqno(2.7)
$$
These transformations are canonical, in the sense that the redefined fields
satisfy the same set of OPEs as the original ones.  We have taken the
opportunity to rescale away some tiresome numerical coefficients at the same
time. In terms of the new fields, which we shall use exclusively from now
on, the BRST operator becomes [9]
$$
Q_B=Q_0+Q_1,\eqno(2.8)
$$
where
$$
\eqalignno{
Q_0&=\oint dz\, c \Big(T^{\rm eff} +T_{\varphi} + T_{\gamma,\beta} + \ft12
T_{c,b} \Big), &(2.9)\cr
Q_1&=\oint dz\, \gamma\Big( (\del\varphi)^3 + 3\alpha\, \del^2\varphi\, \del
\varphi + \ft{19}8 \del^3\varphi +\ft92 \del\varphi\, \beta\, \del\gamma
+\ft32 \alpha\, \del\beta\, \del\gamma\Big),&(2.10)\cr}
$$
with the energy-momentum tensors given by
$$
\eqalignno{
T_\varphi&\equiv -\ft12 (\del\varphi)^2 -\alpha\, \del^2\varphi, &(2.11)\cr
T_{\gamma,\beta}&\equiv -3\, \beta\,\del\gamma -2\, \del\beta\, \gamma,
&(2.12)\cr
T_{c,b}&\equiv -2\, b\, \del c - \del b\, c, &(2.13)\cr
T^{\rm eff} &\equiv -\ft12 \del X^\mu\, \del X^\nu\, \eta_{\mu\nu} -
i a_\mu\, \del^2 X^\mu. &(2.14)\cr}
$$
The BRST operator is graded, with $Q_0^2=Q_1^2=\{Q_0,Q_1\}=0$.

   The physical states of the $W_3$ string have been analysed in considerable
detail.  The early discussions [5,3,10,11] concentrated on physical states
having what one might call ``standard'' ghost structure.  In ordinary
bosonic string theory,  standard ghost structure means that the physical
operators $V(z)$ that create physical states $\ket{\rm phys}=V(0)\ket{0}$
are of the form $V=c\, Y(X)$, where $c$ is the ghost field and $Y(X)$ is
independent of ghosts. ($Y=e^{i p\cdot X}$ for tachyons, $\xi\cdot\del X ^{i
p\cdot X}$ at level 1, {\it etc}.  Note that here, and always in this paper,
$\ket{0}$ denotes the $SL(2,C)$-invariant vacuum.)  For the $W_3$ string,
the analogous notion of states with ``standard'' ghost structure is when the
physical operator $V$ takes the form $V=c\, \del\gamma\, \gamma\,
Y(\varphi,X)$.  The physical-state conditions (2.6) imply that such
operators are physical if
$$
V=c\, \del\gamma\, \gamma\, e^{\mu\varphi}\, Y_\Delta^{\rm eff}(X),\eqno(2.15)
$$
where $Y_\Delta^{\rm eff}(X)$ is a highest-weight operator under $T^{\rm eff}$
with weight $\Delta$, and
$$
\mu=-\ft87\alpha, \ \ \Delta=1;\quad {\rm or} \quad
\mu=-\ft67\alpha, \ \ \Delta=1;\quad {\rm or} \quad
\mu=-\alpha, \ \ \Delta=\ft{15}{16}.\eqno(2.16)
$$
Thus we see that the momentum in the $\varphi$ direction is frozen to  three
discrete values.  From the point of view of the effective spacetime described
by the $X^\mu$ coordinates, these physical states look like two sectors of
ordinary bosonic strings, one having intercept 1, and the other having
intercept
$\ft{15}{16}$ [5,10].

     It turns out that this is by no means the end of the story as far as
physical states are concerned.  There are also physical states with
``non-standard'' ghost structure.  The first example of such a state was
found in [12], in the case of the two-scalar $W_3$ string.  Subsequently,
more physical states with non-standard ghost structure were found in [13].
The main focus of [13] was to examine analogues of the ``ground-ring'' of
ghost-number zero physical operators for the two-scalar $W_3$ string.
However, it was also shown that some (but not all) of the
non-standard ghost structure physical states of the two-scalar $W_3$ string
generalise to physical states of the multi-scalar $W_3$ string [13].  For
example, there is a physical state described by the operator
$$
V=c\, \gamma\, e^{\mu\varphi}\, Y_\Delta^{\rm eff}(X) \eqno(2.17)
$$
with $\mu=-\ft47\alpha$ and $\Delta=\ft12$ [13,14,9].  From the effective
spacetime point of view, this corresponds to a third sector of effective
bosonic string states, with intercept $\ft12$.

     We shall not present a detailed discussion of all the physical states  of
the $W_3$ string here, but instead just give a summary of the final
conclusions.  Further details may be found in [14,9].  Let us first consider
what  have been called ``prime states.''  These are the physical states of
lowest  possible ghost number at a given momentum and level.  From the prime
states, one can  build quartets of physical states by normal ordering with
either or both of the  ``ghost boosters'' $a_\varphi\equiv [Q_B,\varphi]$ and
$a_{{\sss X}^\mu} \equiv  [Q_B,X^\mu]$ [13,14,9].  In addition there are
conjugate quartets [13].  Thus  it suffices to describe the prime states in
order to characterise the entire  physical spectrum.

     The physical prime states divide into four sectors.  The first three of
these  sectors comprise physical states with continuous on-shell spactime
momentum,  and are described by operators of the form [9]
$$
V=c\, U(\varphi,\beta,\gamma)\, Y_\Delta^{\rm eff}(X).\eqno(2.18)
$$
The operator $Y^{\rm eff}_\Delta(X)$ is highest-weight under $T^{\rm eff}$,
with  conformal weight $\Delta$ taking one of the three values 1,
$\ft{15}{16}$, $\ft12$.  $U(\varphi,\beta,\gamma)$ has conformal weight
$h=1-\Delta$ under $T_\varphi+T_{\gamma,\beta}$, and in addition satisfies
$[Q_1,U\}=0$.  For simplicity, we may take the operator $Y^{\rm eff}_\Delta(X)$
to  be tachyonic for now, since the process of constructing excited spacetime
operators is identical to that in ordinary bosonic string theory.  Thus we
may classify the physical states by their level number $\ell$, which
represents the level of ghost and $\varphi$ excitations only.  One finds
that as one goes to higher and higher levels, the same set of three $\Delta$
values 1, $\ft{15}{16}$, $\ft12$ recur repeatedly, with the ghost numbers
of the associated operators $U(\varphi,\beta,\gamma)$ becoming more and more
negative, and the $\varphi$ momentum frozen to more and more positive
values.  The lowest-level examples for $\Delta=1$ and $\Delta=\ft{15}{16}$
are the standard ghost structure $\ell=0$ operators (2.16), and the
lowest-level
example for $\Delta=\ft12$ is the level $\ell=1$ operator (2.17).  Many more
examples may be found in [14,9].

     The fourth sector of physical prime states in the multi-scalar $W_3$
string  comprises states with discrete momentum (in fact $p_\mu=0$ or
$p_\mu=-2a_\mu$) in the  effective spacetime described by $X^\mu$.  Those with
$p_\mu=0$ take the form
$$
V= c\, U_1(\varphi,\beta,\gamma) + U_2(\varphi,\beta,\gamma),\eqno(2.19)
$$
where the operators $U_i(\varphi,\beta,\gamma)$ satisfy the conditions
consequent upon (2.6). The lowest-level examples occur
at $\ell=1$ [13], but a more interesting discrete state is the one at level
$\ell=6$, which is described by the operator [13,9]
$$
D=\Big[ c\,\beta
+\ft14\Big(6\sqrt{2}\, \del \beta\, \gamma
 -3\sqrt{2}\,\beta\,\del\gamma+ 12 \del
\varphi\,  \beta\,  \gamma+ 4\sqrt{2}\, \del
\varphi\,\del\varphi\, + 2 \del^2 \varphi\,\Big)
\Big]e^{\ft27\alpha\varphi}.\eqno(2.20)
$$

     From the pattern of physical states, their frozen $\varphi$-momenta and
ghost numbers, obtained in [14,9], one can see that the level-6 discrete
operator $D$ , and its associated screening current $\oint b\, D$, might be
used to build up each entire tower of physical states in each sector from its
lowest-level member.  Thus the $\Delta=1$ and $\ft{15}{16}$ level $\ell=0$
states (2.15) [5,10], and the $\Delta=\ft12$ level $\ell=1$ state (2.17) [13]
could be viewed as the basic building blocks for all the continuous-momentum
physical states of the multi-scalar $W_3$ string. This idea was first proposed
in [13], where the discrete state $D$ (2.20) was derived as one of the
ground-ring generators of the $W_3$ string.  The fermionic screening current
$\oint b\, D$ was first explicitly presented in [15]; in the simplified
formalism  introduced in [9] that we are using in this paper, it takes the form
$\beta\, e^{\ft27Q\varphi}$ [9] (clearly in general, every discrete operator
(2.19) has an associated screening current $U_1(\varphi,\beta,\gamma)$).
Dealing with  the multiple contour integrals that arise is very cumbersome,
since only certain powers of $D$ give well-defined products.  In practice, it
seems that the easiest way to obtain explicit expressions for higher-level
physical states is by directly solving the physical-state conditions (2.6). The
utility of the screening current $\beta\, e^{\ft27Q\varphi}$ is largely
restricted to a descriptive r\^ole in organising the higher-level physical
spectrum once it is already known by other means, rather than as a tool for
{\it deriving} the full cohomology of the BRST operator.  Some further details
of its action were presented in [16], where its use in
obtaining physical states already found in [14,9] was exhibited.   It has
recently  been shown that there are discrete physical operators at level
$\ell=15$ which are invertible, and are thus guaranteed to give
BRST-non-trivial physical states when normal ordered with any physical
operators [17].  These, by contrast, can be used to {\it derive} the complete
cohomolgy of the $W_3$ string, both in the two-scalar and the multi-scalar
cases [17].

\bigskip
\noindent{\it 2.2 Interactions}
\medskip

     Constructing BRST-invariant scattering amplitudes for the $W_3$ string
remained an outstanding problem for quite a long time.  The difficulty was
that with the states of standard ghost structure known at that time, it was
simply not possible to build multi-point correlation functions in which the
total ghost number of the operators took the correct value ($G_{b,c}=3$ and
$G_{\beta,\gamma}=5$) and the total $\varphi$ momentum took the correct value
($-2\alpha$, where $\alpha$ is the background charge).  One can easily see
this by looking at the form of the standard ghost structure states (2.16).

     The key to building $W_3$ scattering amplitudes was the discovery
[12,13] of physical states in the multi-scalar $W_3$ string with
non-standard ghost structure.  Now, it turns out that both of the above
difficulties are eliminated at a stroke [14,9].  In fact, one builds
BRST-invariant $W_3$-string scattering amplitudes by {\it precisely} the
same techniques as one uses for the ordinary bosonic string [14,9].  In
other words, one simply builds all possible correlation functions of the
four sectors of physical operators of the theory.  Those which achieve the
correct balance of $\varphi$ momentum and the correct overall ghost structure
can be non-zero.  (The use of the ghost boosters $a_\varphi$ and
$a_{{\sss X}^\mu}$ is often important in order to achieve the correct overall
ghost structure.) For three-point functions, this is the end of the story;
since the physical operators have conformal weight zero under the total
energy-momentum tensor, the results are simply constants, independent of the
locations of the three operators on the worldsheet.  For higher $N$-point
functions, it is necessary, as in ordinary bosonic string theory, to make the
replacement  $V(z_i)\longrightarrow\oint dw\, b(w) V(z_i)$ for $(N-3)$
of the physical operators, so as to make spin-1 currents
(screening operators) that can be integrated over the positions $z_i$ to give
an invariant  amplitude [14,9].\footnote{$^\dagger$}{\tenfoot After this
procedure for computing scattering amplitudes for the $W_3$ string was obtained
in [14,9], papers appeared [16] making the incorrect claim that the
prescription in [14,9] was incomplete.   Specifically, the authors of [16]
claimed that unless one augmented the prescription of [14,9] by introducing a
{\it new kind} of amplitude in which a certain screening charge was
included, then one could not construct all the non-zero amplitudes of the $W_3$
string. The claim in [16] is obviously wrong, since the authors simply
rediscovered the screening current $\oint dw\, b(w) D(z)$ where $D(z)$ is given
by (2.20). (As is well known, acting on {\it any}
physical operator with $\oint dw\, b(w)$ gives a screening current.)  Thus
inclusion of the screening charge is just a particular case of the general
procedure of computing $N$-point correlation functions of physical
operators that we described above, where one of the $N>3$ operators
happens to be the discrete physical operator $D$ of (2.20).  Apparently the
authors of [16] were unaware that the screening current was simply $\oint dw\,
b(w) D$, and therefore was already included in the procedure for calculating
scattering amplitudes given in [14,9].}

      We shall not give any detailed scattering-amplitude calculations here;
many examples can be found in [14,9].  The upshot is that  the ``tower'' of
physical  operators in any one of the four sectors described above all behave
equivalently  from the effective spacetime point of view, and essentially can
be
thought  of as equivalent representatives of the same effective-spacetime
physical  state.  Thus one effectively has three Virasoro-like string sectors,
but with  intercepts $\Delta=1$, $\ft{15}{16}$ and $\ft12$, and a fourth sector
of discrete  operators which, having $\Delta=0$ in the effective spacetime,
all behave like representatives of the identity.

     To uncover the pattern of  non-vanishing correlation functions, one must
choose representatives  appropriately from the four sectors (and use
ghost-boosters if necessary) so as to achieve the correct $\varphi$-momentum
balance and overall ghost  structure.  Note that for some amplitudes the fourth
sector, of discrete physical states, can  play an essential r\^ole in achieving
the $\varphi$-momentum balance.  This  is the case for example for the
scattering of four physical states from the  $\Delta=\ft{15}{16}$ sector.  In
an obvious notation $\fpf{\ft{15}{16}} {\ft{15}{16}}{\oint
b\ft{15}{16}}{\ft{15}{16}}$ itself is zero, but
$\fipf{\ft{15}{16}}{\ft{15}{16}}{\oint b\ft{15}{16}}{\oint b D}{\ft{15}{16}}$
is non-zero, where $D$ is the discrete physical operator (2.20).  From the
effective-spacetime point of view, the latter is also a four-point function,
since $D$ has $p_\mu=0$. (It happens that this four-point function was not
originally evaluated in [14,9].  Possibly the authors of [16] mistakenly
thought that this meant that it could not be calculated, giving rise to the
incorrect claims in [16] discussed in the footnote on the previous page.)

     The full set of tree-level scattering amplitudes for the multi-scalar
$W_3$
string turns out to admit a very simple interpretation:   The scattering
amplitudes are exactly what one would get for a critical Virasoro string in
which one tensors the $c=\ft{51}2$ energy-momentum tensor (2.5) with the
$c=\ft12$ energy-momentum tensor for the Ising model.  In fact, if one
associates the $\Delta=1$, $\ft{15}{16}$ and $\ft12$ sectors of the theory with
the identity operator 1, the spin field $\sigma$, and the energy operator
$\epsilon$ of the Ising model, then the $W_3$ scattering amplitudes  reproduce
the fusion rules of the Ising model [14,9].

     In a sense, this conclusion is a somewhat unexciting one.  What one has
learned is that the operators $U(\varphi,\beta,\gamma)$ in (2.18) provide a
representation of the primary fields of the Ising model, realised on the
$(\varphi,\beta,\gamma)$ system.  (From (2.11) and (2.12) one easily sees
that $T_\varphi$ has central charge $\ft{149}2$, and $T_{\gamma,\beta}$ has
central charge $-74$, so their total is indeed the central charge $\ft12$ of
the Ising model.)  Some further aspects were discussed in [18].

     As mentioned in the introduction, if one solves the
physical-state conditions (2.6) for the case of the {\it two-scalar} $W_3$
string ({\it i.e.}\ just one field $X$ in addition to the field $\varphi$), the
spectrum turns out to have many more states than simply those corresponding to
specialising the physical states discussed in section {\it 2.1} to the case of
a single $X$ coordinate [12,13,14,9,19,17].  To put it another way, the
two-scalar $W_3$ string has additional states in its physical spectrum that
do not generalise to the multi-scalar case.  These states have the property
that they cannot be written in a factorised form such as (2.18) or (2.19).
For example, at level $\ell=3$ there is a physical operator $\big[21b\, c\,
\gamma -28 \del\gamma -\ft{16}3 \alpha\, c +12(\alpha\,\del\varphi + a\,
\del X)\,\gamma \big] e^{-\ft37\alpha\varphi +\ft37 a X}$, where
$a=\alpha/\sqrt{3}$ is the background charge for the $X$ coordinate.  For such
states, there is no effective-spacetime interpretation, and the $W_3$ symmetry
evidently acts in a more non-trivial way than it does on the states of the
multi-scalar $W_3$ string.  It may well be, therefore, that the two-scalar
$W_3$ string will provide more insights into the meaning of $W_3$ geometry.

\bigskip
\noindent{\bf 3. Higher-spin generalisations}
\medskip

     A natural generalisation of the above discussion would be to consider a
string theory based on a different $W$ algebra, such as the $W_N$ algebra.  It
is believed that the multi-scalar $W_N$ string would admit an effective
spacetime interpretation as a $c=26-\big[1-\ft{6}{N(N+1)}\big]$ Virasoro-type
string tensored with the $N$'th unitary Virasoro minimal model [5,20].  More
interesting possibilities emerge if one considers a $W$ string with fewer
higher-spin currents, since there will be correspondingly fewer constraints and
therefore a richer physical spectrum.  A case that has been considered recently
is where one has just two currents, namely the energy-momentum tensor and a
primary current of spin $s$ [6].  The BRST operator for such a string can be
obtained by writing an ansatz that generalises (2.8)--(2.14).  In fact the only
changes are to replace (2.10) by
$$
Q_1=\oint dz\, \gamma\, F(\varphi,\beta,\gamma),\eqno(3.1)
$$
where $F(\varphi,\beta,\gamma)$ is a spin-$s$ operator to be solved for, and
$(\beta,\gamma)$ is now a ghost-system for spin $s$, so $\beta$ has spin $s$,
$\gamma$ has spin $(1-s)$, and (2.12) is replaced by
$$
T_{\gamma,\beta}=-s\, \beta\, \del\gamma -(s-1)\, \del\beta\, \gamma.\eqno(3.2)
$$
One can now attempt to solve for operators $F(\varphi,\beta,\gamma)$ which give
a nilpotent BRST charge.

    Solutions for $F(\varphi,\beta,\gamma)$ were found in [6] for the cases of
$s=4$, 5 and 6.  It seems likely that solutions exist for arbitrary $s$, but
their complexity grows rapidly with increasing $s$.  In fact for $s=4$ there
are two distinct solutions, for $s=5$ there is one, and for $s=6$ there four.
Just one solution in each case seems to be associated with a unitary string
theory; it corresponds to $T^{\rm eff}$ having the central charge $c^{\rm eff}=
26 -\ft{2(s-2)}{(s+1)}$ [6].

     We shall discuss only the solutions for $Q_B$ that appear to be associated
with unitary string theories here.  It seems that  in the multi-scalar case the
physical states can again be interpreted as coming from a Virasoro-like string
theory, in this case tensored with a $c=\ft{2(s-2)}{(s+1)}$ minimal model.
This is the central charge of the lowest unitary minimal model of
the $W_{s-1}$ algebra.  The physical states for the models with $s=4$, 5 and 6
were analysed in some detail in [6], and evidence was found
supporting the conjecture that the physical states correspond to those for an
effective Virasoro string tensored with the lowest unitary $W_{s-1}$ minimal
model.  In particular, it was found in [6] that some of the continuous-momentum
physical operators, which take the general form (2.18), correspond to effective
Virasoro operators $Y^{\rm eff}(X)$ with intercepts $\Delta$ that are conjugate
to the conformal weights $h$ of the primary fields of the relevant $W_{s-1}$
minimal model, in the sense that $\Delta=1-h$.  The remaining
continuous-momentum physical operators correspond to intercepts that are
conjugate to weights having the form of a positive integer plus a weight of
a primary field of the minimal model.  This can be easily understood, and was
investigated in detail in [21]:

     The fields of the $W_{s-1}$ minimal model can be viewed as fields of a
Virasoro model with the same central charge.  Primary fields of the $W_{s-1}$
model will also be primary under Virasoro.  In addition, fields that are $W$
secondaries in the $W_{s-1}$ model ({\it i.e.}\ fields that are built by acting
with the negative modes of the higher-spin currents of the $W_{s-1}$ algebra)
will also be Virasoro primaries, with weights that are increased by some
integers relative to those of the $W_{s-1}$ primaries.  Thus what we are seeing
explicitly in these examples is the way in which the set of primary fields of a
given Virasoro model can be more economically described in terms of a smaller
set of primaries of a $W$ minimal model of the same central charge.  When
$s=4$, the $W_3$ minimal model has central charge $c=\ft45$, so there is just a
{\it finite} set of Virasoro primaries in this case.  However, for $s=5$ and
$s=6$ the corresponding $W_4$ and $W_5$ minimal models have central charges
$c=1$ and $c=\ft87$ respectively.  In these cases, an infinite number of
Virasoro primaries are obtained from a finite number of $W$ primaries and their
$W$ descendants.

     We have seen for the examples of $s=4$, 5 and 6 that the lowest unitary
$W_{s-1}$ minimal model, with central charge $c=\ft{2(s-2)}{(s+1)}$, can be
realised in terms of the $(\varphi,\beta,\gamma)$ system, with the primary and
$W$-secondary fields of the model described by the set of operators
$U(\varphi,\beta,\gamma)$  appearing in the physical operators (2.18) of the
theory.  Included amongst the $W_{s-1}$ primary operators are the primary
currents of the $W_{s-1}$ algebra themselves, with spins 3, 4, $\ldots (s-1)$.
In fact they arise as physical operators with zero $\varphi$ momenta, at levels
that can easily be seen to be given by all integers $\ell$ in the interval
$\ft12 s(s-1) + 3\le \ell\le \ft12 s(s+1) -1$.  Thus we obtain an explicit
realisation of $W_{s-1}$ at central charge  $c=\ft{2(s-2)}{(s+1)}$ in terms of
a
scalar field $\varphi$ and the $(\beta,\gamma)$ ghost system with spins
$(s,1-s)$.  Of course one can bosonise the $(\beta,\gamma)$ ghosts, thereby
obtaining a two-scalar realisation of $W_{s-1}$ at the specific value of
central charge given above.  In interesting feature of these rather unusual
realisations is that the OPEs of these primary currents close on the relevant
$W_{s-1}$ algebra modulo the appearance of certain additional null fields on
the
right-hand side [21].  For example, the case $s=4$ gives a two-scalar
realisation at $c=\ft45$ which closes on $W_3$ modulo the appearance of an
additional spin-4 primary null field in the OPE of the spin-3 generator with
itself. Such realisations were studied previously in [22].

%\np
\bigskip
\noindent{\bf 4. Conclusions}
\medskip

     In this paper we have presented a brief review of the current status of
some of the developments in $W$-string theory.  We have concentrated mostly on
the $W_3$ string, since this is the simplest non-trivial example and it already
illustrates many of the new features that result from generalising ordinary
bosonic string theory.  We have also concentrated mostly on the multi-scalar
case, since this enables one to have sufficiently many dimensions that
one can give the theory a reasonably sensible spacetime interpretation.  As we
have seen, the spacetime theory that emerges is essentially that of a
$c=\ft{51}2$ Virasoro string tensored with the Ising model.

     Theories that are perhaps more interesting arise if we consider a $W$
string in which the higher-spin local worldsheet symmetries are relatively
sparse.  In particular, if one considers the case of a $W$ string with just
two local symmetries, generated by currents of spins 2 and $s$, then as we saw
in section 3 the multi-scalar theory is essentially equivalent to that of a
$c=26-\ft{2(s-2)}{(s+1)}$ Virasoro string tensored with the lowest unitary
$W_{s-1}$ minimal model.  The way in which this $W_{s-1}$ symmetry is realised
on the physical states, and the manner in which it enables one to classify the
physical states of the theory, is quite intriguing.

     In some sense the more subtle aspects of the $W$ symmetry are lost in the
multi-scalar $W$ string.  This reflects itself in the way in which the physical
states factorise, as in (2.18), into the product of Virasoro states times
minimal-model fields.  If, however, one works instead with the basic
``Miura''-type realisations of the $W$ algebra, one finds that there are many
more physical states that cannot factorise in this way, suggesting that they
correspond to a more non-trivial realisation of the $W$ symmetry.  For the
$W_3$ string, and the spin-2 plus spin-$s$ strings discussed in section 3,
these basic realisations are in terms of just two scalar fields, $\varphi$ and
$X$.  If one is looking for a non-trivial realisation of the $W$ symmetry, as
opposed to a ``realistic'' spacetime theory, these basic ``pure $W$-gravity''
theories are probably of greater interest.

     We have not touched on several important topics, notably the construction
of a non-critical BRST operator for the $W_3$ string [15,23], which has two
independent realisations of the $W_3$ symmetry, one for the ``Liouville
sector''
and the other for the ``matter sector.''  This is an extremely interesting area
deserving of further study.  Another topic that we have not covered is the
attempt to give a full and rigorous derivation of the cohomology of the BRST
operator.  Progress in this direction has recently been achieved in [19,17].

\bigskip
\noindent{\bf Acknowledgments}
\medskip
     I am very grateful to my collaborators Hong Lu, Bengt Nilsson, Stany
Schrans, Ergin Sezgin, Kelly Stelle, Kris Thielemans, Xujing Wang and Kaiwen
Xu.

\np

\singlespace
\centerline{\bf REFERENCES}
\frenchspacing
\medskip

\item{[1]}A.B. Zamolodchikov, {\sl Teor. Mat. Fiz.} {\bf 65} (1985)
1205; V.A. Fateev and A.B. Zamolodchikov, {\sl Nucl. Phys.} {\bf B280}
(1987) 644; V.A. Fateev and S. Lukyanov, {\sl Int. J. Mod. Phys.} {\bf A3}
(1988) 507.

\item{[2]}A. Bilal and J.-L. Gervais, {\sl Nucl. Phys.} {\bf B314} (1989)
646.

\item{[3]}C.N. Pope, L.J. Romans and K.S. Stelle, {\sl Phys.
Lett.} {\bf B268} (1991) 167; {\sl Phys. Lett.} {\bf B269} (1991) 287.

\item{[4]}L.J.  Romans, {\sl Nucl.  Phys.} {\bf B352} (1991) 829.

\item{[5]}S.R. Das, A. Dhar and S.K. Rama, {\sl Mod. Phys. Lett.}
{\bf A6} (1991) 3055; {\sl Int. J. Mod. Phys.} {\bf A7} (1992) 2295.

\item{[6]}H. Lu, C.N. Pope and X.J. Wang,  ``On higher-spin
generalisations of string theory,''  CTP TAMU-22/93,
hep-th/9304115.

\item{[7]}J. Thierry-Mieg, {\sl Phys. Lett.} {\bf B197} (1987) 368.

\item{[8]}F. Bais, P. Bouwknegt, M. Surridge and K. Schoutens, {\sl
Nucl. Phys.}\ {\bf B304} (1988) 348.

\item{[9]}H. Lu, C.N. Pope, S. Schrans and X.J. Wang,  ``On the spectrum
and scattering of $W_3$ strings,'' CTP TAMU-4/93,
hep-th/9301099, to appear in {\sl Nucl. Phys.} {\bf B}.

\item{[10]}C.N. Pope, L.J. Romans, E. Sezgin and K.S. Stelle,
{\sl Phys. Lett.} {\bf B274} (1992) 298.

\item{[11]}H. Lu, B.E.W. Nilsson, C.N. Pope, K.S. Stelle and P.C. West,
{\sl Int. J. Mod. Phys.} {\bf A8} (1993) 4071.

\item{[12]}S.K. Rama, {\sl Mod.\ Phys.\ Lett.}\ {\bf A6} (1991) 3531.

\item{[13]}C.N. Pope, E. Sezgin, K.S. Stelle and X.J. Wang, {\sl Phys. Lett.}
{\bf B299} (1993) 247.

\item{[14]}H. Lu, C.N. Pope, S. Schrans and X.J. Wang, {\sl Nucl. Phys.} {\bf
B403} (1993) 351.

\item{[15]}M. Bershadsky, W. Lerche, D. Nemeschansky and N.P. Warner, {\sl
Nucl.
Phys.} {\bf B401} (1993) 304.

\item{[16]}M.D. Freeman and P.C. West, ``The covariant scattering and
cohomology of $W_3$ strings,'' KCL-TH-93-2, hep-th/9302114;
``$W_3$ strings, parafermions and the Ising model,'' KCL-93-10,
hep-th/9306134.

\item{[17]}H. Lu, C.N. Pope, X.J. Wang and K.W. Xu, ``The complete cohomology
of the $W_3$ string,'' CTP TAMU-50/93, hep-th/9309041.

\item{[18]}C.M. Hull, ``New realisations of minimal models and the structure
of $W$ strings,'' NSF-ITP-93-65, hep-th/9305098.

\item{[19]}P. Bouwknegt, J. McCarthy and K. Pilch, ``Semi-infinite
cohomology of $W$ algebras,'' USC-93/11, hep-th9302086.

\item{[20]}H. Lu, C.N. Pope, S. Schrans and K.W. Xu,  {\sl Nucl.
Phys.} {\bf B385} (1992) 99.

\item{[21]}H. Lu, C.N. Pope, K. Thielemans and X.J. Wang, ``Higher-spin strings
and $W$ minimal models,''  CTP TAMU-43/93, KUL-TF-93/34, hep-th/9308114, to
appear in {\sl Class. Quantum Grav.}

\item{[22]}E. Bergshoeff, H.J. Boonstra and M. de Roo, ``Realisations of $W_3$
symmetry,''  UG-6/92, hep-th/9209065.

\item{[23]}E. Bergshoeff, A. Sevrin and X. Shen, {\sl Phys. Lett} {\bf B296}
(1992) 95.

\end